 \documentclass[final,3p,times,twocolumn]{elsarticle}
   \usepackage{graphicx}
\usepackage{amssymb}
\usepackage{amsmath}
\journal{Physics Letter B}

\begin{document}

\begin{frontmatter}



\title{Variation after projection calculations for high-spin states}


\author[1]{Zao-Chun Gao\corref{cor1}}
\ead{zcgao@ciae.ac.cn}
\cortext[cor1]{Corresponding author}
\address[1] {China Institute of Atomic Energy, P.O. Box 275 (10), Beijing 102413, PR China}

\begin{abstract}
In this paper, an improvement has been made on the variation after projection (VAP) method, which is crucial in the calculations of high-spin states. It turns out that, the form of the trial VAP wave function with spin $J$ can be simplified by adopting just one projected state rather than previously adopting all $(2J+1)$  angular momentum projected states for each selected reference state, $|\Phi\rangle$. The present calculations show that such simplification still minimizes the angular momentum projected energy up to a very good approximation. In this simplified VAP scheme, one can obtain almost equivalent VAP wave functions starting from different sets of the projected basis states. This clearly shows that a nuclear state can not be identified
with a single intrinsic state, while in the traditional nuclear collective models, an intrinsic state is usually assigned to a well deformed rotational band.
\end{abstract}


\end{frontmatter}

% \linenumbers


Projection is a powerful technique that has long been used in various fields of physics, such as particle physics \cite{Lubeck86}, atomic physics  \cite{Bylicki92} and the physics of condensed matter  \cite{Yang87}. In nuclear physics, the projection method has also been used by many authors to improve the quality of the nuclear wave functions \cite{Hara95,Enyo98,Otsuka01,Bender03,Schmid04,Bender08,gao09,Yao10,Bally14,gao15,tuya17,Robledo18,wang18,Yao20,Shimizu21,Bally21}. This technique removes the spurious part of the trial wave function and makes the projected wave function rather suitable in describing various properties of nuclei and other quantum systems.
 The projected wave functions with different forms have been varied by several authors \cite{Enyo98,Schmid04,Shimizu21,gao15,tuya17,wang18}, so that they can be as close as possible to the corresponding eigenfunctions  obtained by the shell model(SM)  \cite{Caurier05}. Such variation of the projected wave function is generally called as variation after projection (VAP)  \cite{Ring80}. The VAP methods are expected to be applicable in large model space where full shell model calculation can not be performed presently or in the near future.

If the VAP includes the angular momentum projection, it is believed that all the angular momentum projected states should be used to construct the VAP wave function. That means if one uses $n$ reference states to construct a VAP wave function with spin $J$, then the number of included projected states is $n(2J+1)$. In the calculations of high-spin states, this number may be large. Clearly, too many projected states may increase the complexity of the VAP calculation. Actually, large number of projected states may increase the possibility of redundant projected states and may damage the stability of the VAP iteration.

In this Letter, an improvement is made that only $n$ projected states are adopted to construct the VAP wave function with arbitrary spin $J$. This makes the VAP calculations for high-spin states as easy as those for $J=0$ states. For simplicity, the reference states used in the present VAP  are Slater determinants (SD), so that the particle number projections can be omitted.

Let me start with the simplest variation of a SD, $|\Phi\rangle$, i.e., the Hartree-Fock (HF) mean-field approximation. One can vary $|\Phi\rangle$ so that the expectation energy, ${\langle\Phi|\hat H|\Phi\rangle}$, reaches a minimum. However, the minimized HF energy and the corresponding wave function $|\Phi\rangle$ are usually far away from the exact SM ones  \cite{gao09}.

 In the general case, the HF mean-field approximation breaks the symmetries of rotation and reflection. One may project $|\Phi\rangle$ to get a set of  projected states, $P^J_{MK}P^\pi|\Phi\rangle$, where, $P^J_{MK}$ and $P^\pi$ are the angular momentum and parity projection operators, respectively. The state $P^J_{MK}P^\pi|\Phi\rangle$ has good quantum numbers of spin ($J$), the magnetic quantum number in the laboratory frame ($M$) and parity ($\pi$). Notice that, $K$ is not a good quantum number, as will be discussed later.

 In my formalism, the SD $|\Phi\rangle$ can be assumed to be fully symmetry-unrestricted, this makes it convenient to calculate all kinds of states in even-even, odd-odd and odd-mass nuclei in the same way. Additionally, the converged VAP energy can be lower than the one with symmetry-restricted SD. But for the sake of simplicity, the presently used model spaces include only one single harmonic oscillator (HO) shell. The SDs in such model spaces have automatically a good parity. Therefore, I forget about the parity projection and only the angular momentum projection is considered in the present work.

 With given $J$ and $M$, one can project $|\Phi\rangle$ onto $2J+1$ different projected states, $P^J_{MK}|\Phi\rangle$, with $K=-J$, $-J+1$, $\cdots$, $J$.
The nuclear wave function can be expressed in terms of these projected states
\begin{eqnarray}\label{wf}
|\Psi_{JM\alpha}\rangle=\sum_{K=-J}^Jf^{J\alpha}_{K}P^J_{MK}|\Phi\rangle,
\end{eqnarray}
where $\alpha$ is used to label the states with the same $J$ and $M$.
The coefficients $f^{J\alpha}_{K}$ and the corresponding energy $E^J_\alpha$ of Eq. (\ref{wf})
can be determined by solving the Hill-Wheeler (HW)
 equation of order $2J+1$,
\begin{eqnarray}\label{hw}
\sum_{K'=-J}^{J}(H^J_{KK'}-E^J_\alpha N^J_{KK'})f^{J\alpha}_{K'}=0,
\end{eqnarray}
where $H^J_{KK'}=\langle\Phi|\hat H P^J_{KK'}|\Phi\rangle$ and $N^J_{KK'}=\langle\Phi|P^J_{KK'}|\Phi\rangle$. For convenience, I assume $E^J_1\leq E^J_2\leq\cdots\leq E^J_{2J+1}$. The coefficients $f^{J\alpha}_{K'}$ should satisfy the normalization condition.
\begin{eqnarray}\label{norm}
\sum_{K,K'=-J}^{J}f^{J\alpha*}_{K}N^J_{KK'}f^{J\alpha}_{K'}=1.
\end{eqnarray}

In Eq. (\ref{hw}), $f^{J\alpha}_{K}$ is independent of $M$. Since $M$ does not carry any information of physics, it will no longer be discussed and simply be regarded as a fixed quantum number.

To solve Eq. (\ref{hw}), the first step is the diagonalization of $N^J$ and one has
\begin{eqnarray}\label{dn}
\sum_{K'=-J}^J N^J_{KK'} R^k_{K'}=n_kR^k_K,
\end{eqnarray}
 where $n_k  \geq 0$ and $R^k$ with $k=1,2,...2J+1$ are eigenvalues and the corresponding eigenvectors, respectively. Here, I assume $n_1\geq n_2\geq \cdots \geq n_{2J+1}$. Then one can establish a new set of orthonormal basis states, $|\psi_k^J\rangle \,(k=1,2,\cdots,2J+1)$,
 \begin{eqnarray}\label{psi}
 |\psi_k^J\rangle=\frac {1}{\sqrt{n_k}} \sum_{K=-J}^J R^k_K P^J_{MK}|\Phi\rangle,
 \end{eqnarray}
and the HW equation (\ref{hw}) can be transformed into a normal eigenvalue equation
\begin{eqnarray}\label{eigen}
\sum_{k'=1}^{2J+1}\left [\langle \psi_k^J|\hat H|\psi_{k'}^J\rangle-E^J_\alpha \delta_{kk'}\right ]u^{J\alpha}_{k'}=0.
\end{eqnarray}
Actually, the energies in Eq. (\ref{hw}) are obtained by solving Eq. (\ref{eigen}) and the coefficients of the wave functions, $f^{J\alpha}_{K}$, are obtained from $u^{J\alpha}_{k}$, i.e.,
\begin{eqnarray}\label{eigen1}
f^{J\alpha}_{K}=\sum_{k=1}^{2J+1} \frac{R^k_K u^{J\alpha}_k}{\sqrt{n_k}}.
\end{eqnarray}

 The problem is, the $2J+1$ projected states are not orthogonal to one another. Hence in practical calculations, it is possible that some $n_k$ values can be very tiny or even zero. Consequently, the corresponding $|\psi_k^J\rangle$ basis states may not be precise enough to guarantee the stability of the calculated energies and wave functions. To avoid this trouble, the $|\psi_k^J\rangle$ states with tiny $n_k$ values should be abandoned. This can be done by setting a cutoff parameter $\epsilon>0$ and only those $|\psi_k^J\rangle$ states with $n_k>\epsilon$ are taken to form the nuclear wave function (see, e.g., Ref.  \cite{Zdunczuk07}).

Such basis cutoff works if $|\Phi\rangle$ remains unchanged. However, if one tries to vary $|\Phi\rangle$ in attempt to get the best nuclear wave function $|\Psi_{JM\alpha}\rangle$, in other words, to perform the VAP calculation, new trouble arises due to the changes of $n_k$ values during VAP iteration. When a fixed cutoff parameter $\epsilon$ is used, the number of selected  $|\psi_k^J\rangle$ states likely changes as VAP iteration goes on. This will destroy the smoothness of the energy as a function of the variational parameters. On the other hand, if one fixes the number of selected $|\psi_k^J\rangle$ states, then some of them might have too small $n_k$ values at certain VAP iteration. Certainly, both treatments of the $|\psi_k^J\rangle$ selection do not guarantee the stability of the VAP iteration. These problems are exemplified in the figures included in the supplemental material.

 One can imagine that the safest VAP calculation might be the one that only $|\psi_1^J\rangle$ with the largest $n_1$ is taken. But $|\psi_1^J\rangle$ still includes several projected states $P^J_{MK}|\Phi\rangle$. Sometimes, some of them might have very tiny norms, $\langle\Phi|P^J_{KK}|\Phi\rangle$, or two projected states might be almost the same. For instance, $P^J_{MK}|\Phi\rangle=(-1)^{J-K}P^J_{M,-K}|\Phi\rangle$ when $|\Phi\rangle$ is time-even  \cite{gao02}. These problems may still affect the precision of the calculated nuclear wave functions. Finally, it looks the only way of ensuring the VAP stability is that only one projected state $P^J_{MK}|\Phi\rangle$ is taken. Then, the nuclear wave function can be simplified as
\begin{eqnarray}\label{wf_k}
|\Psi_{JM}(K)\rangle=\frac{P^J_{MK}|\Phi\rangle}{\sqrt{\langle \Phi|P^J_{KK}|\Phi\rangle}},
\end{eqnarray}
 where, $K$ can be randomly chosen but should satisfy $\langle \Phi|P^J_{KK}|\Phi\rangle\neq 0$. $K$ is put into the bracket since it is not a good quantum number.

Now, there have two forms of nuclear wave functions, Eq. (\ref{wf}) and Eq. (\ref{wf_k}), to be optimized.
One may expect Eq. (\ref{wf}) is better than Eq. (\ref{wf_k}) if they have the same $|\Phi\rangle$. Actually,  in almost all previous VAP works  \cite{Schmid04,Shimizu21,tuya17,wang18}, the former Eq. (\ref{wf}) was taken.  However, if one do the VAP calculations with these two different forms of nuclear wave functions, independently, the results are quite interesting.

\begin{figure}
  \includegraphics[width=3.3in]{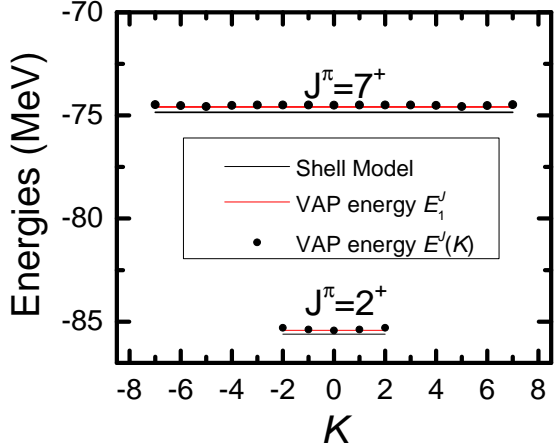}
  \caption{ (Color online)  Calculated VAP energies, $E^J_1$ (red lines) and $E^J(K)$(filled dots), for the yrast $2^+$ and $7^+$ states in $^{24}$Mg, with wave functions Eq.(\ref{wf})  and Eq.(\ref{wf_k}), respectively. The corresponding exact shell model energies are shown as black lines. The USDB interaction is adopted.}\label{fig1}
\end{figure}

As a first example of such calculations, the USDB interaction  \cite{usdb} defined in the $sd$ shell is used. The lowest $J^\pi=2^+$ and $7^+$ energies in $^{24}$Mg are calculated. The energy of Eq. (\ref{wf_k}) is written as
\begin{eqnarray}\label{ejk}
E^J(K)=\langle \Psi_{JM}(K)|\hat H|\Psi_{JM}(K)\rangle.
\end{eqnarray}
For each $K$, $E^J(K)$ is minimized independently and one gets $2J+1$ converged $E^J(K)$ energies at spin $J$. The VAP method of Ref.  \cite{tuya17} is applied to minimize $E^J_1$ in Eq.(\ref{hw}) and $E^J(K)$ in Eq(\ref{ejk}), respectively.
The results are shown in Fig. \ref{fig1} (see more details in the supplemental material). It is seen that the $E^J_1$ energies are very close to the exact ones calculated by the full SM which has been discussed in Ref.  \cite{tuya17}. Surprisingly, one can also see that all the $E^J(K)$ energies are almost the same as $E^J_1$ energies for both $J^\pi=2^+$ and $7^+$ without exception. Correspondingly, the $2J+1$ converged wave functions $|\Psi_{JM}(K)\rangle$ for $E^J(K)$ are obtained. If $K$ is a good quantum number, these $|\Psi_{JM}(K)\rangle$ states (with common fixed $J$ and $M$) should be orthogonal to each other. This is clearly not possible because it is known the norm matrix, $N^J$, in Eq. (2) is usually not diagonal. On the contrary, it is expected they are the same state since all $E^J(K)$ energies converge to the same level, which wave function should be unique, thus one should have
\begin{eqnarray}\label{overlap}
|\langle\Psi_{JM}(K)|\Psi_{JM}(K')\rangle|\approx 1
\end{eqnarray}
for arbitrary $K$ and $K'$. All possible overlaps, $|\langle\Psi_{JM}(K)|\Psi_{JM}(K')\rangle|$, for both $J^\pi=2^+$ and $7^+$ states have been calculated. Most of them are above 0.99 and the worst overlap is still over 0.98. This clearly tells us that one may arbitrarily select a $K$ number and use Eq. (\ref{wf_k}) to do the VAP calculation without losing good approximation.

Before performing further calculations, it is necessary to understand the reason why $E^J(K)$ energies are almost the same as $E^J_1$. At a given spin $J$, one can assume that the nucleus is rotating. Once one has the optimized SD $|\Phi\rangle$ for Eq. (\ref{wf}), $E^J_1$ is then obtained from the Eq. (\ref{hw}). This also determines the orientation of rotational axis relative to $|\Phi\rangle$ for $E^J_1$. Since $E^J_1$ is the lowest, the moment of inertia along this rotational axis should be the largest. On the other hand, one can first fix the orientation of the rotational axis relative to $|\Phi\rangle$ and then do the energy minimization, which is actually done by taking Eq.(\ref{wf_k}). It is expected that the moment of inertia along this fixed rotational axis can also reach the same maximum as that for $E^J_1$, and $E^J(K)$ converges to $E^J_1$, simultaneously.

  Thus it is shown that the VAP calculations with wave functions in Eq. (\ref{wf}) and Eq. (\ref{wf_k}) are almost equivalent, which seems to be a universal phenomenon. If this is true, Eq. (\ref{wf_k}) can be used to simplify the VAP calculation and make the VAP iteration equally stable at any spin. This is crucial in extending the present VAP calculations to arbitrary high-spin states.

 If one takes Eq. (\ref{wf_k}) as the nuclear wave function, then it can be fully obtained from the $K$-projected state, $P^K|\Phi\rangle$. Here $P^K=\frac{1}{2\pi}\int_0^{2\pi}e^{i(K-\hat J_z)\phi}d\phi$ is the $K$-projection operator. When two different reference states, $|\Phi\rangle$ and $|\Phi'\rangle$, are taken, one can do the $K$-projection on them, and get $P^K|\Phi\rangle$ and $P^{K'}|\Phi'\rangle$ with $K\neq K'$. These $K$-projected states are strictly orthogonal to one another, but their total angular momentum ($J$) projected states can be almost identical according to Eq. (\ref{overlap}).

So far, there is only one SD in Eq. (\ref{wf_k}). To improve the VAP approximation, Eq. (\ref{wf_k}) should be generalized by including more SDs, and the new form of VAP wave function can be written as,
\begin{eqnarray}\label{wf_nk}
|\Psi_{JM\alpha}(K)\rangle=\sum_{i=1}^n f^{J\alpha}_{i}P^J_{MK}|\Phi_i\rangle,
\end{eqnarray}
where, $n$ is the number of included $|\Phi_i\rangle$ SDs. $f^{J\alpha}_i$ is determined by the following Hill-Wheeler equation,
\begin{eqnarray}\label{hw_k}
\sum_{i'=1}^{n}(H^J_{ii'}-E^J_\alpha N^J_{ii'})f^{J\alpha}_{i'}=0,
\end{eqnarray}
where $H^J_{ii'}=\langle\Phi_i|\hat H P^J_{KK}|\Phi_{i'}\rangle$ and $N^J_{ii'}=\langle\Phi_i| P^J_{KK}|\Phi_{i'}\rangle$.

One may expect Eq. (\ref{wf_nk}) can also be used to describe the non-yrast states. To check the validity of this assumption, here a newly developed VAP algorithm  \cite{wang18} is applied, in which the low-lying state wave functions with the same spin and parity can be varied on the same footing. This can be safely realized by minimizing the sum of the corresponding low-lying energies, $\sum_{\alpha=1}^m E_\alpha^J$. Here $m$ is the number of calculated states.

It should be reminded there are two more potential problems with the projected basis in Eq. (\ref{wf_nk}), which may seriously affect the stability of the VAP calculation. The one is the norm, $N^J_{ii}$, could become very tiny as the VAP iteration proceeds. The other is the possibility of large overlaps among the projected basis states. Both problems may damage the precision of the calculated energies and the corresponding wave functions, and may cause the collapse of the VAP iteration. Fortunately, such troubles do not appear in the calculations of Fig. \ref{fig1}. However, to ensure the stability of VAP iteration, in the following calculations, two constraint terms are attached to the energy sum, and the final minimized quantity is,
 \begin{eqnarray}\label{2}
 Q=\sum_{\alpha=1}^m E^J_\alpha+\chi_1\sum_{i=1}^n\frac 1{N^J_{ii}}+\frac{\chi_2}2\sum_{\substack{i,j=1\\i\neq j}}^n\frac{N^J_{ij}N^J_{ji}}{N^J_{ii}N^J_{jj}},
 \end{eqnarray}
where the second term tends to push the norms, $N^J_{ii}$, to large values, and the third term tends to guide the projected basis states to be orthogonal to one another. The values of the last two terms should be as small as possible provided that the VAP iteration is stable. So the parameters $\chi_1 \geq 0$ and $\chi_2\geq 0$ should not be too large. In all the following calculations, I use $\chi_1=10^{-4}$MeV and $\chi_2=1$MeV. This makes the included projected states almost orthogonal to one another after the VAP calculation converges. It turns out that both quantities of the last two terms in Eq. (\ref{2}) are within 100keV in the present calculations  (see more detailed calculations in the supplemental material).

\begin{figure*}
  \includegraphics[width=6.6in]{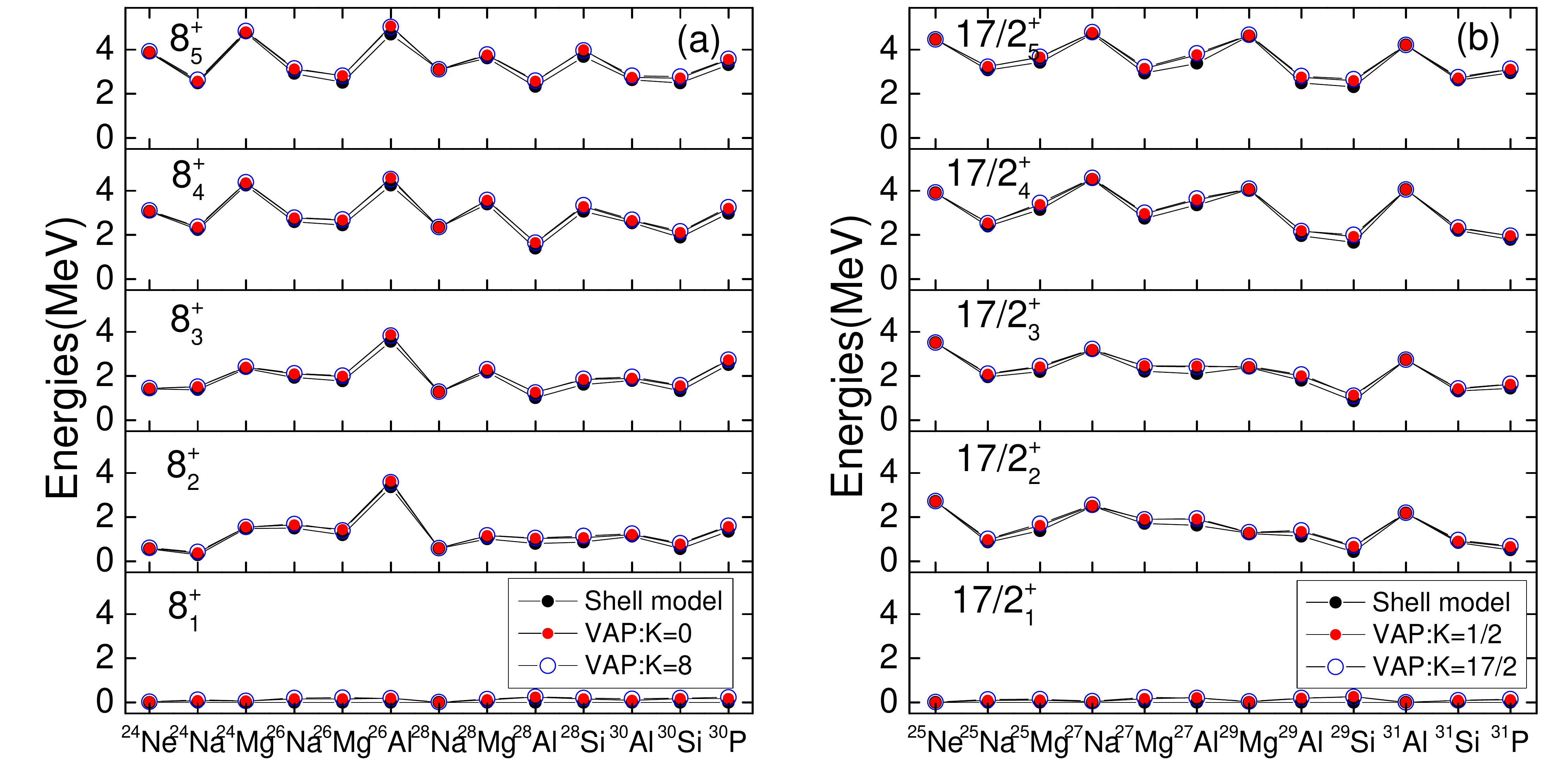}
  \caption{ (Color online)   Calculated lowest five $J^\pi=8^+$ eneriges in some $sd$ shell nuclei with the Shell model and the VAP, respectively. The five panels from low to high show the energies from the yrast $8^+_1$ ones to the forth excited $8^+_5$ ones, respectively. VAP:$K=0$ and VAP:$K=8$ refer to the results of the VAP calculations with  $|\Psi_{JM\alpha}(K=0)\rangle$ and $|\Psi_{JM\alpha}(K=8)\rangle$, respectively. For each nucleus, all calculated energies are shifted by the same quantity so that the yrast SM enery is zero (see the lowest panel). The USDB interaction is adopted.}\label{spin}
\end{figure*}

 Now, let us come to the calculations for the non-yrast states. In the second example, the USDB interaction is adopted again and the results are shown in Fig. \ref{spin}. Here, the wave functions $|\Psi_{JM\alpha}(K)\rangle$ [see Eq. (\ref{wf_nk})] with $n=10$ are taken to be varied. The quantitiy of Eq. (\ref{2}) including the sum of the lowest five energies ($m=5$) is minimized. To show the generality of Eq. (\ref{wf_nk}), the calculations include even-even, odd-odd and odd-mass nuclei. For even-even and odd-odd nuclei, the calculated spin is taken to be $J=8$, and for the odd-mass ones, $J=17/2$. As we learned from the first example, the $K$ number in $|\Psi_{JM\alpha}(K)\rangle$ may be arbitrarily chosen within the range $|K|\leq J$. For the $J=8$ case, $K=0$ and $K=8$ are chosen, and the corresponding VAP energies are denoted by VAP:$K=0$ (red dots) and VAP:$K=8$ (blue circles), respectively in Fig. \ref{spin}a. Similar calculations are performed for the $J=17/2$ states but $K=1/2$ and $K=17/2$ are chosen and the corresponding energies are shown in Fig. \ref{spin}b. Let us first look at the Fig. \ref{spin}a. All the calculated energies of VAP:$K=8$ perfectly coincide with those of VAP:$K=0$. This implies that the results of VAP:$K=0$ and VAP:$K=8$ are the same. One can imagine VAP results with other $K$ should also be the same as VAP:$K=0$ (or VAP:$K=8$). Comparing with the full shell model, the energies of both VAP:$K=0$ and VAP:$K=8$ are very close to the same exact ones. Thus it is expected that both $|\Psi_{JM\alpha}(K=0)\rangle$ and $|\Psi_{JM\alpha}(K=8)\rangle$ are close to the same shell model wave function for each calculated state. This means $|\Psi_{JM\alpha}(K=0)\rangle$ and $|\Psi_{JM\alpha}(K=8)\rangle$ are almost the same. Indeed, for all calculated states in Fig. \ref{spin}a, it is found that
\begin{eqnarray}\label{overlap1}
|\langle\Psi_{JM\alpha}(K=0)|\Psi_{JM\alpha}(K=8)\rangle|\geq 0.95.
\end{eqnarray}
Actually, most of such overlaps are above $0.98$. For the $J=17/2$ states, the equivalence of $|\Psi_{JM\alpha}(K=1/2)\rangle$ and $|\Psi_{JM\alpha}(K=17/2)\rangle$ are also numerically confirmed.

 One can understand that, for an arbitrary selected $K$ with $|K|\leq J$, if the number of selected $|\Phi_i\rangle$ SDs, $n$, is large enough, then the space spanned by the $\hat P^J_{MK}|\Phi_i\rangle$ states may fully cover the whole $J$-scheme shell model configuration space without including other projected basis states, $\hat P^J_{MK'}|\Phi_i\rangle$ with $K'\neq K$. Together with the good approximation of the present calculations, it seems that there is no need to consider the complicate $K$-mixing in all VAP calculations.

\begin{figure}
\centering
  \includegraphics[width=3.3in]{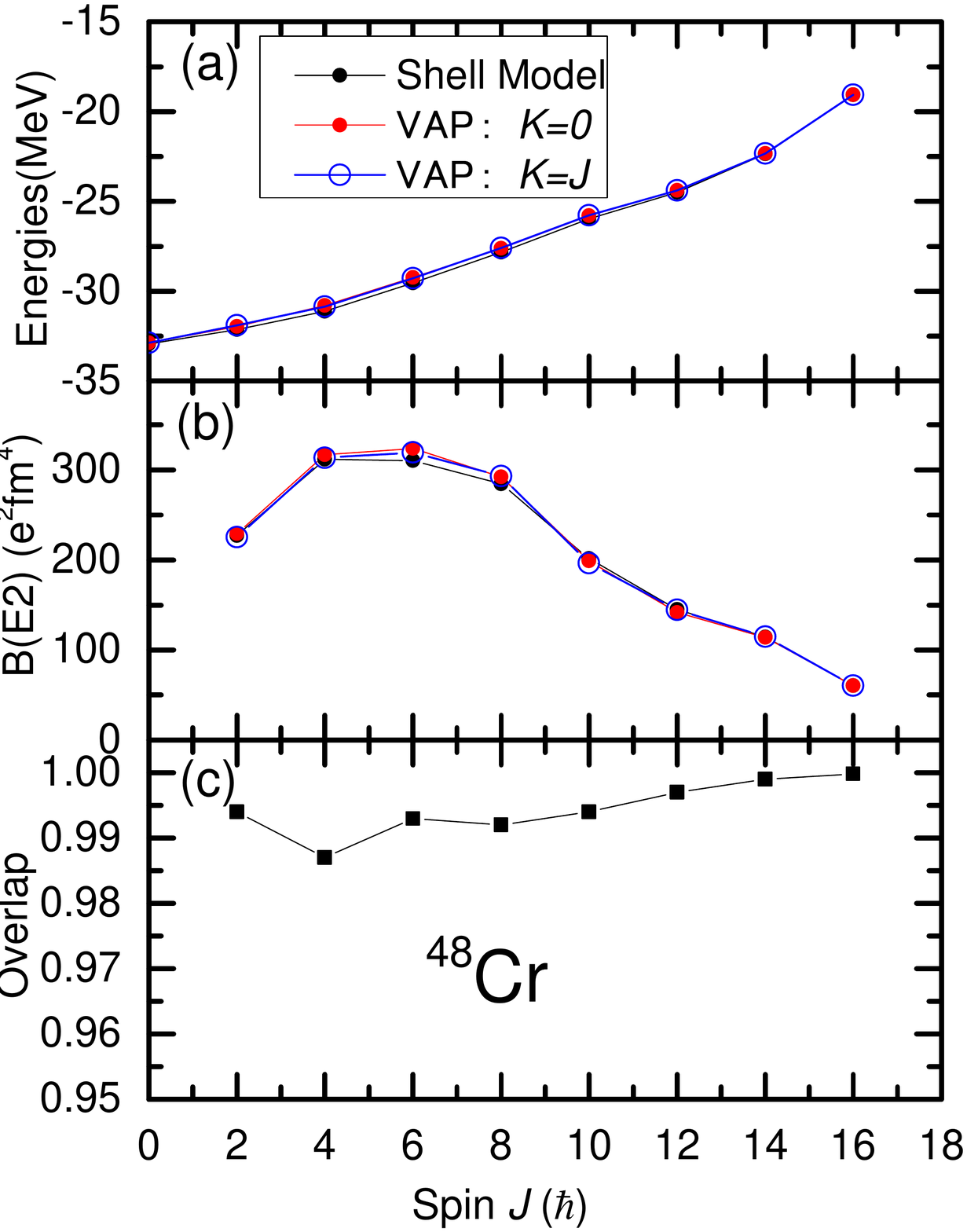}
 \caption{ (Color online)  Calculated results of the yrast states in $^{48}$Cr with the VAP and the shell model. (a) The calculated energies. (b) The B(E2) values from the wave functions corresponding to (a). (c) The values of the overlap, $|\langle\Psi_{JM1}(K=0)|\Psi_{JM1}(K=J)\rangle|$. The KB3 interaction is adopted.}\label{Cr48}
\end{figure}

As a more practical application, the high-spin states in $^{48}$Cr are calculated, which has been studied by the shell model  \cite{Caurier95}, the projected shell model  \cite{Hara99}, and the projected configuration interaction(PCI) method \cite{Gao11}. Here the KB3 interaction  \cite{kb3} is taken, as has been used in Ref.s \cite{Caurier95,Gao11}. This time, the wave functions $|\Psi_{JM\alpha}(K=0)\rangle$ and $|\Psi_{JM\alpha}(K=J)\rangle$ are used to minimize the yrast energies($\alpha=1$) in $^{48}$Cr, respectively. In practical calculations, the $Q$ quantity of Eq. (\ref{2}) with m=1 and n=5 is minimized. The calculated results have been shown in Fig. \ref{Cr48}. From  Fig. \ref{Cr48}(a), the VAP energies with both $K=0$ and $K=J$ are very close to the SM ones. The B(E2) values calculated with the wave functions corresponding to Fig. \ref{Cr48}(a) are shown in Fig. \ref{Cr48}(b). All the B(E2) values are bunched up tightly. The B(E2) values with wave functions $|\Psi_{JM1}(K=0)\rangle$ and $|\Psi_{JM1}(K=J)\rangle$ are almost coincide with each other. This again implies $|\Psi_{JM1}(K=0)\rangle$ and $|\Psi_{JM1}(K=J)\rangle$ are almost the same. The overlap, $|\langle\Psi_{JM1}(K=0)|\Psi_{JM1}(K=J)\rangle|$, as a function of spin $J$ has been shown in Fig. \ref{Cr48}(c). Indeed, the overlaps for all spins are above $98\%$.

 Therefore, all the present calculations have clearly shown that the VAP wave function can be expressed in terms of the projected states, $P^J_{MK}|\Phi_i\rangle$, with the same $K$ values. Without the need of performing the $K$-mixing, the present VAP method can be easily extended to high-spin states. If one prefers to use HFB vacua rather than taking the SDs in VAP, it is likely such extension may still works. However, the reference states can not be uniquely determined in the VAP calculation at a given spin. So, the intrinsic state can not be identified for a single nuclear state. In the traditional nuclear collective models, an intrinsic state is usually assigned to a well deformed rotational band. Nevertheless, if one wants to use the simple, but limited, picture where a single intrinsic state is associated with the ground-state wave function of a rotational band within a VAP scheme, a possible way is to take a common reference state and generate all angular momentum
($J$) projected states with the same $K$ values from it, then minimize the sum of the projected energies of the band members. In that case, the intrinsic state and the $K$ value might be determined. The details of such work will be published in the future.

\textbf{Acknowledgements} This work is supported by the National Natural Science Foundation of China under Grant Nos. 11975314,11575290,
and by the Continuous Basic Scientific Research Project Nos. WDJC-2019-13, BJ20002501.

%% References
%% Usage of  \cite is as follows:
%%    \cite{key}          ==>>  [#]
%%    \cite[chap. 2]{key} ==>>  [#, chap. 2]
%%    \citet{key}         ==>>  Author [#]


%\bibliographystyle{model1a-num-names}



%\begin{thebibliography}{00}

\end{document}